\documentclass[12pt]{article}
\usepackage{amssymb}
\usepackage{graphicx}

\begin{document}

\begin{titlepage}
\begin{center}

\vspace*{10mm}

{\LARGE\bf
Dynamical Electroweak Symmetry Breaking in String Models with D-branes}

\vspace*{20mm}

{\large
Noriaki Kitazawa
}
\vspace{6mm}

{\it
Department of Physics, Tokyo Metropolitan University,\\
Hachioji, Tokyo 192-0397, Japan\\
e-mail: kitazawa@phys.metro-u.ac.jp
}

\vspace*{15mm}

\begin{abstract}
The possibility of dynamical electroweak symmetry breaking
 by strong coupling gauge interaction in models with D-branes
 in String Theory is examined.
Instead of elementary scalar Higgs doublet fields,
 the gauge symmetry with strong coupling (technicolor) is introduced.
As the first step of this direction,
 a toy model, which is not fully realistic,
 is concretely analyzed in some detail.
The model consists of D-branes and anti-D-branes
 at orbifold singularities in $(T^2 \times T^2 \times T^2)/{\bf Z}_3$
 which preserves supersymmetry.
Supersymmetry is broken through the brane supersymmetry breaking.
It is pointed out
 that the problem of large $S$ parameter
 in dynamical electroweak symmetry breaking scenario
 may be solved by natural existence of kinetic term mixings between
 hypercharge U$(1)$ gauge boson and massive anomalous U$(1)$ gauge bosons.
The problems to be solved toward constructing more realistic models
 are clarified in the analysis.
\end{abstract}

\end{center}
\end{titlepage}

\section{Introduction}
\label{introduction}

The quest for natural mechanism of electroweak symmetry breaking
 is one of the most important subject in high energy physics.
Supersymmetric extension of the standard model
 \cite{Haber:1993wf,Nilles:1995ci,Baer:1995tb,Bagger:1996ka,
 Castano:1993ri,Drees:1995hj}
 is the most popular scenario
 because of smooth transition into the grand unification
 with radiative electroweak symmetry breaking \cite{Inoue:1982pi}.
Instead of supersymmetry,
 some global symmetries also can act the role
 to stabilize the electroweak symmetry breaking scale
 \cite{ArkaniHamed:2001nc,ArkaniHamed:2002qy}.
The other popular scenario is assuming extra space dimensions
 \cite{Antoniadis:1990ew,ArkaniHamed:1998rs,Randall:1999ee}
 by which Planck scale of gravity is reduced to the scale
 of electroweak symmetry breaking.
The existence of extra dimensions also makes possible
 to break gauge symmetry by boundary conditions,
 which is utilized to ``unify'' Higgs doublet fields and gauge fields
 \cite{Fairlie:1979at,Manton:1979kb,Hosotani:1983xw}
 or to break electroweak symmetry without Higgs doublet fields
 \cite{Csaki:2003dt}.
The possibility of dynamical electroweak symmetry breaking
 \cite{Weinberg:1975gm,Weinberg:1979bn,Susskind:1978ms,
 Miransky:1988xi,Nambu:1989jt,Marciano:1989xd,Bardeen:1989ds}
 is still pursued in spite of that precision electroweak data
 put this possibility at a disadvantage
 (see Ref.\cite{Belyaev:2008yj} for a review).

In phenomenological efforts in String Theory,
 many works have focused on
 providing appropriate boundary conditions at high energies
 for successful supersymmetric extensions of the standard model
 or supersymmetric grand unified theories in field theory
 without theoretical inconsistencies.
Flux compactifications for moduli stabilizations
 can give warped extra dimensions and natural electroweak symmetry breaking
 \cite{Giddings:2001yu},
 but technical difficulties do not allow explicit quantitative analysis
 in String Theory.
Although some scenarios utilizing ``stringy'' effects,
 Higgs doublet fields as tachyons in D-brane recombinations
 \cite{Aldazabal:2000dg,Cremades:2002cs},
 symmetry breaking by string loop corrections
 \cite{Antoniadis:2000tq,Kitazawa:2006if},
 for example, have been proposed,
 technical difficulties prevent to construct concrete calculable models.

In this paper
 we focus on models with D-branes in type II string theory.
There are two types of models:
 models with D-branes intersecting in compact space
 \cite{Bachas:1995ik,Blumenhagen:2000wh,Angelantonj:2000hi,
 Aldazabal:2000dg,Aldazabal:2000cn,Ibanez:2001nd,Blumenhagen:2001te,
 Cvetic:2001tj,Cvetic:2001nr,Kokorelis:2002zz}
 and models with D-branes at singularities in compact space
 \cite{Aldazabal:2000sa}.
In both types of models
 there is a tendency for the number of Higgs doublet fields to be many.
It was shown that
 the number of (massless) Higgs doublet fields
 in the first type of models should be larger than three
 \cite{Higaki:2005ie},
 if we consider simple factorized toroidal compactifications
 $T^2 \times T^2 \times T^2$ with some orbifold or orientifold projections.
(This is not the case,
 if Higgs doublet fields come from stretched open string
 and originally massive \cite{Ibanez:2001nd}.
 More general compactifications
 also change the situation \cite{Anastasopoulos:2006da}.
 Introduction of the compositeness may also change the situation
 \cite{Kitazawa:2004ed,Kitazawa:2004hz,Kitazawa:2004nf}.)
The similar situation happens in the second type of models.

In case with extra Higgs doublet fields,
 it is very difficult to escape from too large FCNC
 to be consistent with experiments.
If all the Higgs doublet fields get vacuum expectation values,
 the Yukawa coupling matrices of massive neutral scalar fields
 are not diagonal in mass eigenstates of quarks and leptons,
 and these neutral scalar fields mediate large FCNC.
It is very unnatural
 to assume very large masses for these neutral scalar fields.
Even if only one Higgs doublet fields gets vacuum expectation value,
 the situation is the same without assuming very large masses
 for extra Higgs doublet fields.
This fact does not depend on
 whether low-energy supersymmetry exists or not.

In case of low string scale with large compact dimensions
 for radiative electroweak symmetry breaking by string loop corrections,
 massive string modes associated with extra Higgs doublet fields
 also mediate large FCNC.
It can be shown that
 the four-fermion processes mediated by Higgs exchanges
 are also dressed by string form factors
 in the similar way that is shown in Ref. \cite{Antoniadis:2000jv}
 on the processes mediated by gauge boson exchanges.
The lower bound on the string scale $2 \sim 3$ TeV
 is obtained in Ref.\cite{Antoniadis:2000jv},
 which is critical for radiative electroweak symmetry breaking.
In the models of the first type
 there are further additional sources of FCNC
 by the world-sheet instanton effect
 which is necessary to generate Yukawa couplings
 for the masses of quarks and leptons.
A strong constraint on the string scale
 to be larger than about $10^3$ TeV is obtained
 in Ref.\cite{Abel:2004rp}
 assuming simple compactifications which make models calculable.
Therefore, radiative breaking is not the possibility
 for the first type of models.
Astrophysical constraints on the sizes of extra dimensions
 (we will discuss this constraint in section \ref{toy-model} in detail)
 almost excluded the possibility of the string scale of a few TeV
 and of radiative breaking scenario even in the second type of models
 \footnote{
 There are two types of massless scalar modes
  on a D-brane at a singularity:
  one is associated with D-brane moduli
  and the other is associated with Wilson line \cite{Kitazawa:2008tb}.
 The scale of mass correction for the first type is typically
  one order lower than the string scale
  due to the contributions from twisted closed strings.
 Though the scale of mass correction for the second type
  is originally governed by the compactification scale
  (and can be smaller than the string scale),
  we have to use orbifold projection to extract scalar fields
  in non-adjoint representations (Higgs doublet fields).
 Then the scale of mass correction
  is again one order lower than the string scale,
  because the contributions from twisted closed strings dominate.
 The mass corrections to the massless scalar modes
  from the open strings between different D-branes have not investigated yet.
 }.

It looks that
 the safest way to proceed
 is to find the models with minimal number of Higgs doublet fields
 with low-energy supersymmetry (and its breaking),
 and to provide appropriate boundary conditions for
 standard radiative electroweak symmetry breaking
 taking the string scale very high.
In fact there are many efforts in this direction with moduli stabilization
 (for a resent effort, see Ref.\cite{Conlon:2008wa}, for example).
There is, however,
 a possibility of dynamical electroweak symmetry breaking
 by some strong gauge interaction which has not seriously examined yet.
It is well known that
 this scenario, technicolor, is strongly constrained
 by the precise experimental measurements of the corrections
 to self-energies of electroweak gauge bosons.
Though this problem of large $S$ parameter is fatal
 for simple technicolor dynamics like QCD scaling-up,
 it has been proposed that certain non-trivial dynamics
 do not cause the problem
 \cite{Sundrum:1991rf,Appelquist:1998xf,Hong:2006si}.
It is also pointed out that 
 the existence of massive vector fields
 which mix with hypercharge gauge boson in kinetic term
 (kinetic term mixing) may solve the problem,
 even if the dynamics is simple \cite{Holdom:1990xp,Kitazawa:1995nb}.
We focus on the second possibility,
 since massive anomalous U$(1)$ gauge fields
 which can have kinetic term mixings with hypercharge gauge field
 naturally emerge in models with D-branes.
The aim of this paper is not to propose a realistic model,
 but to examine the possibility by analyzing a simple concrete toy model,
 though it is non-realistic.
The model is based on D-branes at orbifold singularities
 with a supersymmetric $(T^2 \times T^2 \times T^2)/{\bf Z}_3$
 compactification.
Supersymmetry is broken at the string scale
 by introducing anti-D-branes, namely by the brane supersymmetry breaking
 \cite{Sugimoto:1999tx,Antoniadis:1999xk,Angelantonj:1999jh,
       Aldazabal:1999jr,Angelantonj:1999ms},
 and it is assumed that
 all the originally massless scalar modes obtain masses
 of the order of the string scale
 through the tree-level closed string exchanges
 \cite{Kitazawa:2008tb}.
We also assume that gauginos obtain masses of the order of the string scale,
 though some bulk supersymmetry breaking would have to be introduced
 \cite{Antoniadis:2004qn,Antoniadis:2005sd}.
\footnote{
 For model building with different philosophy using AdS/CFT correspondence,
 see Ref.\cite{Carone:2007md}, for example.}

This paper is organized as follows.
In section \ref{Dynamical-EW-br}
 the scenario of dynamical electroweak symmetry breaking in field theory
 is briefly reviewed.
The problem of large $S$ parameter is also reviewed.
In section \ref{Technicolor-on-brane}
 a toy model is constructed and investigated.
The anomaly structure and masses of U$(1)$ gauge fields
 are investigated in detail.
The contributions to $S$ parameter
 through the anomalous U$(1)$ gauge boson exchanges
 are quantitatively estimated,
 and the possibility of dynamical electroweak symmetry breaking
 in String Theory is examined.
In section \ref{conclusions}
 some general comments and conclusions are presented.

\section{Dynamical electroweak symmetry breaking}
\label{Dynamical-EW-br}

We very briefly review technicolor theory \cite{Farhi:1980xs}
 and its problems with precision electroweak data.

Consider the standard model without Higgs doublet field.
The chiral symmetry in quark sector would be spontaneously broken
 by the quark pair condensates due to the strong coupling effect of QCD.
Since electroweak gauge symmetry would be also broken at the same time,
 the mass of the weak bosons would be of the order of 10 MeV
 corresponding to the QCD scale $\Lambda_{\rm QCD} \sim 100$ MeV
 where QCD coupling becomes order one: $\alpha_s \sim 1$.
Though electroweak gauge symmetry would be broken,
 leptons would not obtain masses because of no mediation
 of the breaking to lepton sector.

Introduce additional QCD-like gauge symmetry, SU$(N_{\rm TC})$,
 with additional left-handed fermions (``techni-fermions'')
\begin{eqnarray}
 \left( \begin{array}{c} N_L \\ E_L \end{array} \right)
  &\sim& (N_{\rm TC}, 2)_{0},\\
 (N_R)^c &\sim& (N_{\rm TC}^*, 1)_{-1/2},\\
 (E_R)^c &\sim& (N_{\rm TC}^*, 1)_{+1/2}
\end{eqnarray}
 under SU$(N_{\rm TC}) \times$SU$(2)_L \times$U$(1)_Y$.
All the gauge anomaly is canceled out
 ($N_{\rm TC}$ should be even for SU$(2)_L$ global anomaly cancellation).
Suppose SU$(N_{\rm TC})$, namely technicolor, becomes strong
 at the scale $\Lambda_{\rm TC} \sim 1$ TeV, we expect the condensates
\begin{equation}
 \langle {\bar N_R} N_L \rangle
 = \langle {\bar E_R} E_L \rangle
 \ne 0
\end{equation} 
 by the criteria of maximal remaining gauge symmetry.
This causes electroweak symmetry breaking
 and weak bosons obtain masses of the order of 100 GeV.
The masses of ordinary fermions
 can be produced through four-fermion interactions like
\begin{equation}
 {\cal L}_{\rm f-mass}
  = - {1 \over {M_{uE}^2}} \left( {\bar q}_L^i u_R \right)
                           \left( {\bar N}_R L_{Li} \right)
    - {1 \over {M_{uN}^2}} \left( {\bar q}_L^i u_R \right)
                           \epsilon_{ij}
                           \left( {\bar L}_L^j E_R \right)
    + {\rm h.c.}
\end{equation}
 for the up quark, where $q_L = (u_L \ d_L)^T$, $L_L = (N_L \ E_L)^T$
 and $M_{uE}$ and $M_{uN}$ are scales of the physics
 which mediate electroweak symmetry breaking to the up quark.
The flavor physics has direct connection
 to the origin of these kinds of four-fermion interactions
 among ordinary fermions and techni-fermions.
This is the scenario of (one-doublet) technicolor
 for natural electroweak symmetry breaking
 with dimensional transmutation.

Precision electroweak measurements give constraints on
 vacuum polarizations of weak bosons, $W^\pm$ and $Z^0$ and photon,
 or $W_{1,2,3}$ and $B$ corresponding to SU$(2)_L$ and U$(1)_Y$.
At low energies,
 where we can neglect all the higher derivative terms in the effective action,
 it can be shown that only three parameters are required
 to describe the new physics contributions beyond the standard model
 by virtue of symmetry and renormalization
 \cite{Peskin:1990zt,Peskin:1991sw,Altarelli:1990zd,Altarelli:1991fk}.
Those three parameters are defined as
\begin{eqnarray}
 S &=& - 16 \pi \Pi'_{3Y}(0),\\
 T &=& {{4\pi} \over {s^2 c^2 m_Z^2}}
       \left[ \Pi_{11}(0) - \Pi_{33}(0) \right],\\
 U &=& 16 \pi \left[ \Pi'_{11}(0) - \Pi'_{33}(0) \right],
\end{eqnarray}
 where $s = \sin \theta_W$ and $c = \cos \theta_W$
 with Weinberg angle $\theta_W$,
 and the vacuum polarizations with no Lorentz indices are defined as
\begin{equation}
 \Pi_{AB}^{\mu\nu}(q)
  = \left( g^{\mu\nu} - {{q^\mu q^\nu} \over {q^2}} \right) \Pi_{AB}(q^2),
 \quad
 \Pi'_{AB}(q^2) \equiv {{d \Pi_{AB}(q^2)} \over {d q^2}}
\end{equation}
 with $A,B = 1,2,3,Y$ corresponding to $W_{1,2,3}$ and $B$.
These vacuum polarizations are defined
 so that they include unknown new physics effects only
 (the {\it known} contributions from the standard model are subtracted).
The vacuum polarizations at zero momentum
 describe the corrections to the masses,
 and the derivative of the vacuum polarizations at zero momentum
 describe the corrections to kinetic terms.
We can understand that
 $S$ parameter describes kinetic term mixing between $W_3$ and $B$
 through new physics beyond the standard model.

Since technicolor and techni-fermions are new physics,
 they are constrained through the above three parameters.
The contribution to $S$ parameter can be naively estimated
 by the calculation of techni-fermion one-loop diagrams
 neglecting strong coupling technicolor effects:
\begin{equation}
 S = {{N_{\rm TC}} \over {6\pi}} N_D,
\label{S-techni}
\end{equation}
 where $N_D$ is the number of SU$(2)_L$ doublets of techni-fermions
 ($N_D = 1$ in the present case).
There are estimates for QCD-like technicolor
 including the effects of strong coupling dynamics
 using the technique of scaling up experimental data of real QCD phenomena,
 and the results support that the above simple estimate is rather good.
The experimental constraint, assuming Higgs mass 600 GeV and $S>0$
 (which is appropriate for QCD-like technicolor dynamics),
 is given as $S < 0.09$ at 95\% CL \cite{Amsler:2008zzb},
 and one can conclude that the technicolor scenario
 with QCD-like technicolor dynamics has already been excluded.
 \footnote{
 Some simple models passing this constraint
 have been proposed in Refs.
 \cite{Sannino:2004qp,Dietrich:2005jn,Foadi:2007ue}.}

It should be mentioned that
 experimental constraints on $T$ parameter
 also make the situation of technicolor scenario unfavorable,
 if the physics of ordinary fermion mass generation is considered.
Since top quark is much heavier than bottom quark,
 ``non-oblique corrections'',
 which can not be parametrized by the above three parameters,
 tend to become large as well as $T$ parameter.
This is another difficulty of technicolor scenario,
 which is strongly related with the physics of the mediation of
 electroweak symmetry breaking to quarks and leptons
 (namely, the origin of flavor physics).
In this paper
 we concentrate on $S$ parameter problem
 leaving flavor problems for future researches.

\section{Technicolor in D-brane models}
\label{Technicolor-on-brane}

In the following subsections,
 we construct a toy model using D-branes at orbifold singularities,
 investigate the anomaly structure and masses of anomalous U$(1)$'s,
 and estimate their contributions to $S$ parameter. 

\subsection{A toy model}
\label{toy-model}

We consider a simple compact space of
 $(T^2 \times T^2 \times T^2)/{\bf Z}_3$
 with orbifold projection vector $v = (1/3,1/3,-2/3)$.
There are three orbifold singularities in each torus,
 and they can be specified as I${}_i$, II${}_i$ and III${}_i$
 for torus $i=1,2,3$.
There are 27 singular points in total in this compact space,
 and we can specify one of them as (I${}_1$, II${}_2$, III${}_3$),
 for example.
The model includes D$3$-branes, D$7_i$-branes and their anti-branes,
 where a D$7_i$-brane means a D$7$-brane filling compact space
 except for $i$-th torus.

First, we construct a local model at (I${}_1$, I${}_2$, I${}_3$)
 with brane supersymmetry breaking.
We would like to have an asymptotically free gauge symmetry
 (assuming all scalar fields are decoupled)
 in addition to the gauge symmetry of the standard model.
Introduce D$3$, D$7_1$, D$7_2$, D$7_3$ and $\overline{{\rm D}7}_1$ branes
 with the following transformation matrices on Chan-Paton factors
 under ${\bf Z}_3$ transformation:
\begin{eqnarray}
 \gamma_3 &=&
  {\rm diag}( {\bf 1}_3, \alpha {\bf 1}_2, \alpha^2 {\bf 1}_1 ),
\label{ours-D3} \\
 \gamma_{7_1} &=& \gamma_{7_2} = \gamma_{7_3} =
  {\rm diag}( 0, 0, \alpha^2 {\bf 1}_1 ),
\label{ours-D7} \\
 \gamma_{{\bar 7}_1} &=&
  {\rm diag}( {\bf 1}_3, 0, 0 ),
\label{ours-anti-D7}
\end{eqnarray}
 where ${\bf 1}_a$ means an $a \times a$ unit matrix
 and $\alpha = \exp(i 2 \pi /3)$.
The twisted Ramond-Ramond (R-R) tadpoles are canceled
 by satisfying the condition
\begin{equation}
 3 {\rm Tr} \ \gamma_3 + \sum_{i=1,2,3} {\rm Tr} \ \gamma_{7_i}
                       - \sum_{i=1,2,3} {\rm Tr} \ \gamma_{{\bar 7}_i} = 0.
\label{RR-condition}
\end{equation}
The total gauge symmetry is 
 U$(3) \times$U$(2) \times$U$(1)$ on D$3$-brane,
 U$(1)'_i$ on D$7_i$-branes ($i=1,2,3$),
 and U$(3)$ on $\overline{{\rm D}7}_1$-brane.
The color SU$(3)_c$ and weak SU$(2)_L$ are identified
 as SU$(3)$ in U$(3) =$ SU$(3) \times$U$(1)_3$
 and SU$(2)$ in U$(2) =$ SU$(2) \times$U$(1)_2$ on D3-branes, respectively.
The hypercharge U$(1)_Y$ is defined as
\begin{equation}
 {Y \over 2} \equiv - \left({1 \over 3} Q_3 + {1 \over 2} Q_2 + Q \right),
\end{equation}
 where $Q_3$, $Q_2$ and $Q$ denote the charges of U$(1)$ symmetries
 on D$3$-brane.
This is a simple realization of non-anomalous hypercharge gauge symmetry
 in many possibilities \cite{Anastasopoulos:2006da}.
In the following we will see that
 this simple realization can not give realistic lepton spectrum,
 because three ``higgsinos'' remain after decoupling of all the scalars.
Since the aim of this paper is
 to examine the possibility of technicolor with anomalous U$(1)$'s
 using a simple toy model, we accept this unrealistic situation.
The technicolor gauge symmetry SU$(3)_{TC}$
 is identified as SU$(3)$ in U$(3) =$ SU$(3) \times$U$(1)_{TC}$ 
 on $\overline{{\rm D}7}_1$-brane.

The value of gauge coupling at the string scale
 depends on the compactification scale in general.
From Dirac-Born-Infeld action
 (with $\kappa_{10}^2 = (2\pi)^7 (\alpha')^4/2$ in type IIB theory)
 we obtain gauge couplings for each D-brane as
\begin{eqnarray}
 g_{D3} &=& \sqrt{2 \pi g_s},\\
 g_{D7_1} = g_{\overline{D7}_1}
          &=& \sqrt{2 \pi g_s}
              \left( {{2 \pi \sqrt{\alpha'}} \over {2 \pi r}} \right)^2,\\
 g_{D7_2} = g_{D7_3}
          &=& \sqrt{2 \pi g_s}
              \left( {{2 \pi \sqrt{\alpha'}} \over {2 \pi r}} \right)
              \left( {{2 \pi \sqrt{\alpha'}} \over {2 \pi R}} \right),
\end{eqnarray}
 where $g_s = e^{\langle \phi \rangle}$ is the string coupling
 and $R$ and $r$ are the compactification radii
 for the first torus and second and third tori, respectively
 (the radii of second and third tori are taken to be the same,
 for simplicity). 
We assume $g_{D3} = \sqrt{2 \pi g_s} \simeq 1$
 and do not discuss the possibility to reproduce smaller gauge couplings
 of SU$(2)_L$ and U$(1)_Y$ than that of SU$(3)_c$.
Furthermore,
 we assume $r \simeq \sqrt{\alpha'}$
 for strong coupling technicolor at low energies,
 and assume large $R$ for low string scale of the order of 10 TeV
 to have light anomalous U$(1)$ gauge bosons.
Since the relation between reduced Planck scale $M_P^*$ and
 the string scale $M_s^* \equiv 1/2 \pi \sqrt{\alpha'}$ is given by
\begin{equation}
 M_s^* = \sqrt{{g_s} \over {4\pi}}
         \left( {{2 \pi \sqrt{\alpha'}} \over {2 \pi R}} \right)
         \left( {{2 \pi \sqrt{\alpha'}} \over {2 \pi r}} \right)^2
         M_P^*,
\end{equation}
 we have $R \simeq 0.02 \ \mu m \simeq 5 \ {\rm eV}^{-1}$
 which is not yet excluded by the supernova SN1987a constraint
 $R < 0.066 \ \mu m$ \cite{Hanhart:2001fx}
 \footnote{
 Here, we do not consider stronger constraints from astrophysics
 assuming substantial branching ratios of decays of Kaluza-Klein
 gravitons to photon \cite{Hannestad:2001xi,Hannestad:2003yd},
 which is model dependent.
 }.
Six-dimensional gravitational scale is given by
\begin{equation}
 M_P^{\rm 6D} = \sqrt{M_P^* M_s^*}
                \left( {{2 \pi \sqrt{\alpha'}} \over {2 \pi R}} \right)^{1/2}
              \simeq 30 \ \rm{TeV},
\end{equation}
 which is not yet excluded by the constraint $M_P^{\rm 6D} > 14$ TeV
 \cite{Amsler:2008zzb}.
The gauge couplings on D$7_2$ and D$7_3$ branes are very small
 of the order of $10^{-7}$.

Though twisted tadpole cancellation, eq.(\ref{RR-condition}),
 guarantee the anomaly cancellation for gauge symmetry on D$3$-brane,
 gauge symmetries on D$7$-branes are anomalous
 without canceling twisted R-R tadpoles at all the singular points
 in their world volumes.
After all,
 twisted R-R tadpoles at all the 27 singularities should be canceled.
Table \ref{brane-conf}
 shows D-branes and anti-D-branes
 which concern twisted R-R tadpole cancellation at each 27 singularity.
We do not consider untwisted R-R tadpole cancellation
 (cancellation of mixed gravitational anomalies) and moduli stabilization,
 since they would not be very essential for the results in this paper,
 though moduli stabilizations might be necessary
 to prevent pair annihilations between branes and anti-branes.
Many D-branes and anti-D-branes are introduced
 with the following transformation matrices under ${\bf Z}_3$ transformation.
\begin{eqnarray}
 \gamma_{3'} &=& \gamma_{3''} =
  {\rm diag}( {\bf 1}_3, \alpha {\bf 1}_2, \alpha^2 {\bf 1}_1 ),\\
 \gamma_{3^{(a)}} &=&
  {\rm diag}( {\bf 1}_1, 0, 0 ), \quad a=1 \cdots 6,\\
 \gamma_{{\bar 3}^{(a)}} &=&
  {\rm diag}( 0, 0, \alpha^2 {\bf 1}_1 ), \quad a=1 \cdots 6,\\
 \gamma_{7_1'} &=& \gamma_{7_1''} =
  {\rm diag}( 0, 0, \alpha^2 {\bf 1}_1 ),\\
 \gamma_{{\bar 7}_2'} &=& \gamma_{{\bar 7}_2''} =
  {\rm diag}( 0, 0, \alpha^2 {\bf 1}_2 ),\\
 \gamma_{7_3'} &=& \gamma_{7_3''} =
  {\rm diag}( 0, 0, \alpha^2 {\bf 1}_1 ).
\end{eqnarray}
Most additional D-branes introduce additional U$(1)$ gauge symmetries,
 except for D$3'$, D$3''$, $\overline{{\rm D}7}_2'$
 and $\overline{{\rm D}7}_2''$ branes.
Two copies of the standard model massless field contents
 are realized with each D$3'$ and D$3''$ brane,
 which have communication with our world on D$3$-brane
 at (I${}_1$, I${}_2$, I${}_3$) singularity
 only through massive (heavier than the string scale 10 TeV)
 open string modes and closed string modes (including gravity). 

All the massless fermion modes
 at (I${}_1$, I${}_2$, I${}_3$) singularity
 are listed in Table \ref{massless-modes}.
All the non-Abelian gauge anomalies are canceled.
We assume that
 all the massless scalar modes obtain masses of the order of the string scale
 and decouple from low energies
 through closed string tree-level effects (or open-string one-loop effects)
 in this configuration of brane supersymmetry breaking
 \cite{Kitazawa:2008tb}.
It is naively assumed that
 all the gauginos also obtain masses of the order of the string scale
 and decouple.
Since gaugino Majorana mass
 is protected by approximate (discrete) R symmetry,
 to obtain large gaugino masses,
 introduction of bulk supersymmetry breaking \`a la Sherk-Schwarz
 would be required to appropriately break (discrete) $R$ symmetry
 \cite{Antoniadis:2004qn,Antoniadis:2005sd}.
Technicolor gauge interaction is asymptotically free
 with the coefficient of beta function as
 $b_{\rm TC} = -13/3$ ($b_{\rm TC} = -7/3$ with techni-gluino).
The number of SU$(2)_L$ doublets in techni-fermions is one,
 but the hypercharge assignment is different from that in one-doublet model
 which has been reviewed in the previous section.
This is due to the fact that we have to have three ``higgsinos''
 in this simple hypercharge realization on a D$3$-brane,
 and techni-fermions are necessary to cancel the anomaly of U$(1)_Y$.
The anomaly structure of 31 U$(1)$ gauge symmetry, including U$(1)_Y$,
 is discussed in the next subsection. 

While technicolor running coupling becomes large below the string scale,
 the coefficient of beta function is smaller
 in absolute value than that of QCD.
If these gauge couplings are equal at the string scale,
 the scale of technicolor condensation is lower than QCD scale.
Here, we simply assume large threshold corrections
 (Ref.\cite{Antoniadis:1999ge,Bachas:1996zt,Bachas:1998kr,
            Ghilencea:2002ff,Anastasopoulos:2006hn} for type I theory,
  Ref.\cite{Lust:2003ky,Akerblom:2007np,Benakli:2008ub}
            for intersecting D-brane models,
  and Ref.\cite{Conlon:2009xf,Conlon:2009kt}
            for models with D-branes at singularities)
 by which technicolor coupling is about two times larger than QCD coupling
 at the string scale of the order of 10 TeV (100\% correction).
Then technicolor coupling
 becomes strong at the energy scale of the order of 100 GeV.
This may not be so unreasonable.
Because technicolor lives on D$7$-brane and QCD lives on D$3$-brane,
 only technicolor gauge coupling receives corrections by Kaluza-Klein states.
It also has been pointed out in Ref.\cite{Conlon:2009xf} that
 the effect of winding modes can be significant
 in case of large compactifications.
Another possibility is
 taking compactification scale $r^2$ about two times smaller
 than the string scale. 
However, more natural realizations of strong coupling technicolor
 should be pursued in realistic models.
One possibility is to realize technicolor on D$3$-branes
 and standard model gauge symmetry on D$7$-branes,
 though the realization of hypercharge becomes rather non-trivial.

The problem to look for
 the pattern of tech-fermion condensates,
 namely the problem of vacuum alignment,
 is a difficult issue in this kind of system with complicated interactions.
Usually, it is assumed that the pattern with maximal symmetry is preferred.
Here, we further assume that
 the only two fermions which can be at the same place in compact space
 can condensate.
One preferred pattern is
\begin{eqnarray}
 \langle \Psi_L^{I_3=-1/2} \Psi_E \rangle
 &=& \langle \Psi_L^{I_3=+1/2} \Psi_{7_2} \rangle \ne 0,
\nonumber\\
 \langle \Psi'_L{}^{I_3=-1/2} \Psi_E' \rangle
 &=& \langle \Psi'_L{}^{I_3=+1/2} \Psi_{7'_3} \rangle \ne 0,
\nonumber\\
 \langle \Psi''_L{}^{I_3=-1/2} \Psi_E'' \rangle
 &=& \langle \Psi''_L{}^{I_3=+1/2} \Psi_{7''_3} \rangle \ne 0,
\nonumber\\
 \langle \Psi_{{\bar 7}_2'}^{I_3=-1/2} \Psi_{7_1}^{i=1} \rangle
 &=& \langle \Psi_{{\bar 7}_2'}^{I_3=+1/2} \Psi_{7_1}^{i=2} \rangle \ne 0,
\nonumber\\
 \langle \Psi_{{\bar 7}_2''}^{I_3=-1/2} \Psi_{7_1}^{i=3} \rangle
 &=& \langle \Psi_{{\bar 7}_2''}^{I_3=+1/2} \Psi_{7_3} \rangle \ne 0.
\label{condensates}
\end{eqnarray}
The first two condensates trigger the electroweak symmetry breaking
 in very similar way in one-doublet model in previous section.
The condensates of second and third lines
 result ``electroweak symmetry breaking'' in two mirror worlds.

\begin{table}[t]
\caption{D-brane configuration for twisted R-R tadpole cancellations.}
\label{brane-conf}
\begin{center}
\begin{tabular}{lllllll}
\hline \\
(I${}_1$, I${}_2$, I${}_3$) & \hspace{12pt} &
 D$3$ & D$7_1$ & D$7_2$ & D$7_3$ & $\overline{{\rm D}7}_1$ \\
(II${}_1$, I${}_2$, I${}_3$) & \hspace{12pt} &
 $\overline{{\rm D}3}^{(1)}$ & D$7_1'$ & D$7_2$ & D$7_3$ & \\
(III${}_1$, I${}_2$, I${}_3$) & \hspace{12pt} &
 $\overline{{\rm D}3}^{(2)}$ & D$7_1''$ &	 D$7_2$ & D$7_3$ & \\
(I${}_1$, II${}_2$, I${}_3$) & \hspace{12pt} &
 D$3^{(1)}$ & D$7_1$ & $\overline{{\rm D}7}_2'$ & D$7_3$ &
  $\overline{{\rm D}7}_1$ \\
(I${}_1$, III${}_2$, I${}_3$) & \hspace{12pt} &
 D$3^{(2)}$ & D$7_1$ & $\overline{{\rm D}7}_2''$ & D$7_3$ &
  $\overline{{\rm D}7}_1$ \\
(II${}_1$, II${}_2$, I${}_3$) & \hspace{12pt} &
  & D$7_1'$ & $\overline{{\rm D}7}_2'$ & D$7_3$ & \\
(II${}_1$, III${}_2$, I${}_3$) & \hspace{12pt} &
  & D$7_1'$ & $\overline{{\rm D}7}_2''$ & D$7_3$ & \\
(III${}_1$, II${}_2$, I${}_3$) & \hspace{12pt} &
  & D$7_1''$ & $\overline{{\rm D}7}_2'$ & D$7_3$ & \\
(III${}_1$, III${}_2$, I${}_3$) & \hspace{12pt} &
  & D$7_1''$ & $\overline{{\rm D}7}_2''$ & D$7_3$ & \\
\\ \hline \\
(I${}_1$, I${}_2$, II${}_3$) & \hspace{12pt} &
 D$3'$ & D$7_1$ & D$7_2$ & D$7_3'$ & $\overline{{\rm D}7}_1$ \\
(II${}_1$, I${}_2$, II${}_3$) & \hspace{12pt} &
 $\overline{{\rm D}3}^{(3)}$ & D$7_1'$ & D$7_2$ & D$7_3'$ & \\
(III${}_1$, I${}_2$, II${}_3$) & \hspace{12pt} &
 $\overline{{\rm D}3}^{(4)}$ & D$7_1''$ &	 D$7_2$ & D$7_3'$ & \\
(I${}_1$, II${}_2$, II${}_3$) & \hspace{12pt} &
 D$3^{(3)}$ & D$7_1$ & $\overline{{\rm D}7}_2'$ & D$7_3'$ &
  $\overline{{\rm D}7}_1$ \\
(I${}_1$, III${}_2$, II${}_3$) & \hspace{12pt} &
 D$3^{(4)}$ & D$7_1$ & $\overline{{\rm D}7}_2''$ & D$7_3'$ &
  $\overline{{\rm D}7}_1$ \\
(II${}_1$, II${}_2$, II${}_3$) & \hspace{12pt} &
  & D$7_1'$ & $\overline{{\rm D}7}_2'$ & D$7_3'$ & \\
(II${}_1$, III${}_2$, II${}_3$) & \hspace{12pt} &
  & D$7_1'$ & $\overline{{\rm D}7}_2''$ & D$7_3'$ & \\
(III${}_1$, II${}_2$, II${}_3$) & \hspace{12pt} &
  & D$7_1''$ & $\overline{{\rm D}7}_2'$ & D$7_3'$ & \\
(III${}_1$, III${}_2$, II${}_3$) & \hspace{12pt} &
  & D$7_1''$ & $\overline{{\rm D}7}_2''$ & D$7_3'$ & \\
\\ \hline \\
(I${}_1$, I${}_2$, III${}_3$) & \hspace{12pt} &
 D$3''$ & D$7_1$ & D$7_2$ & D$7_3''$ & $\overline{{\rm D}7}_1$ \\
(II${}_1$, I${}_2$, III${}_3$) & \hspace{12pt} &
 $\overline{{\rm D}3}^{(5)}$ & D$7_1'$ & D$7_2$ & D$7_3''$ & \\
(III${}_1$, I${}_2$, III${}_3$) & \hspace{12pt} &
 $\overline{{\rm D}3}^{(6)}$ & D$7_1''$ &	 D$7_2$ & D$7_3''$ & \\
(I${}_1$, II${}_2$, III${}_3$) & \hspace{12pt} &
 D$3^{(5)}$ & D$7_1$ & $\overline{{\rm D}7}_2'$ & D$7_3''$ &
  $\overline{{\rm D}7}_1$ \\
(I${}_1$, III${}_2$, III${}_3$) & \hspace{12pt} &
 D$3^{(6)}$ & D$7_1$ & $\overline{{\rm D}7}_2''$ & D$7_3''$ &
  $\overline{{\rm D}7}_1$ \\
(II${}_1$, II${}_2$, III${}_3$) & \hspace{12pt} &
  & D$7_1'$ & $\overline{{\rm D}7}_2'$ & D$7_3''$ & \\
(II${}_1$, III${}_2$, III${}_3$) & \hspace{12pt} &
  & D$7_1'$ & $\overline{{\rm D}7}_2''$ & D$7_3''$ & \\
(III${}_1$, II${}_2$, III${}_3$) & \hspace{12pt} &
  & D$7_1''$ & $\overline{{\rm D}7}_2'$ & D$7_3''$ & \\
(III${}_1$, III${}_2$, III${}_3$) & \hspace{12pt} &
  & D$7_1''$ & $\overline{{\rm D}7}_2''$ & D$7_3''$ & \\
\\ \hline
\end{tabular}
\end{center}
\end{table}
\begin{table}[t]
\caption{Massless fermion modes at (I${}_1$, I${}_2$, I${}_3$) singularity.
 The gauge symmetries on additional D-branes
  are considered as global symmetries in this table.
 The gauge symmetries on D$7_i$-brane
  are described as U$(1)'_i$ with $i=1,2,3$.
 The contents above the horizontal line
  are associated with open strings between two D-branes of our world,
  and those below the horizontal line
  are associated with open strings
  between our D-branes and additional D-branes.
}
\label{massless-modes}
\begin{center}
\begin{tabular}{cccccccccc}
U$(3)_{\rm TC}$ & U$(3)_c$ & U$(2)_L$ & U$(1)$ &
 U$(1)'_1$ & U$(1)'_2$ & U$(1)'_3$ &  & U$(1)_Y$ &  \\
\hline\hline
$1$ & $3^*$ & $1$ & $+1$ & $0$ & $0$ & $0$ & $\times 3$ &
 $-2/3$ & $u_R^c$ \\
$1$ & $3$ & $2^*$ & $0$ & $0$ & $0$ & $0$ & $\times 3$ &
 $+1/6$ & $q_L$ \\
$1$ & $1$ & $2$ & $-1$ & $0$ & $0$ & $0$ & $\times 3$ &
 $+1/2$ & ``${\tilde h}$'' \\
$1$ & $3^*$ & $1$ & $0$ & $+1$ & $0$ & $0$ & &
 $+1/3$ & $d_R^c$ \\
$1$ & $1$ & $2$ & $0$ & $-1$ & $0$ & $0$ & &
 $-1/2$ & $l_L$ \\
$1$ & $3^*$ & $1$ & $0$ & $0$ & $+1$ & $0$ & &
 $+1/3$ & $d_R^c$ \\
$1$ & $1$ & $2$ & $0$ & $0$ & $-1$ & $0$ & &
 $-1/2$ & $l_L$ \\
$1$ & $3^*$ & $1$ & $0$ & $0$ & $0$ & $+1$ & &
 $+1/3$ & $d_R^c$ \\
$1$ & $1$ & $2$ & $0$ & $0$ & $0$ & $-1$ & &
 $-1/2$ & $l_L$ \\
$3^*$ & $1$ & $2$ & $0$ & $0$ & $0$ & $0$ & &
 $-1/2$ & $\Psi_L^{I_3}$ \\
$3$ & $1$ & $1$ & $-1$ & $0$ & $0$ & $0$ & &
 $+1$ & $\Psi_E$ \\
$3$ & $1$ & $1$ & $0$ & $-1$ & $0$ & $0$ & $\times 3$ &
 $0$ & $\Psi_{7_1}^i$ \\
$3$ & $1$ & $1$ & $0$ & $0$ & $-1$ & $0$ & &
 $0$ & $\Psi_{7_2}$ \\
$3$ & $1$ & $1$ & $0$ & $0$ & $0$ & $-1$ & &
 $0$ & $\Psi_{7_3}$ \\
\hline
$3^*$ & $1$ & $1$ & $0$ & $0$ & $0$ & $0$ & $\times 2$ &
 $0$ & $\Psi_{{\bar 7}_2'}^{I_3}$ \\
$3^*$ & $1$ & $1$ & $0$ & $0$ & $0$ & $0$ & $\times 2$ &
 $0$ & $\Psi_{{\bar 7}_2''}^{I_3}$ \\
$3$ & $1$ & $1$ & $0$ & $0$ & $0$ & $0$ & &
 $0$ & $\Psi_{7_3'}$ \\
$3$ & $1$ & $1$ & $0$ & $0$ & $0$ & $0$ & &
 $0$ & $\Psi_{7_3''}$ \\
$3^*$ & $1$ & $1$ & $0$ & $0$ & $0$ & $0$ & $\times 2$ &
 $0$ & $\Psi'_L{}^{I_3}$ \\
$3$ & $1$ & $1$ & $0$ & $0$ & $0$ & $0$ & &
 $0$ & $\Psi_E'$ \\
$3^*$ & $1$ & $1$ & $0$ & $0$ & $0$ & $0$ & $\times 2$ &
 $0$ & $\Psi''_L{}^{I_3}$ \\
$3$ & $1$ & $1$ & $0$ & $0$ & $0$ & $0$ & &
 $0$ & $\Psi_E''$ \\
$1$ & $1$ & $1$ & $0$ & $+1$ & $0$ & $0$ & $\times 12$ &
 $0$ & \\
$1$ & $1$ & $1$ & $0$ & $-1$ & $0$ & $0$ & $\times 4$ &
 $0$ & \\
$1$ & $1$ & $1$ & $0$ & $0$ & $+1$ & $0$ & $\times 6$ &
 $0$ & \\
$1$ & $1$ & $1$ & $0$ & $0$ & $-1$ & $0$ & $\times 4$ &
 $0$ & \\
$1$ & $1$ & $1$ & $0$ & $0$ & $0$ & $+1$ & $\times 2$ &
 $0$ & \\
\hline\hline
\end{tabular}
\end{center}
\end{table}

\subsection{Anomalous U$(1)$ gauge bosons}
\label{anomalous-U(1)}

There are 31 U$(1)$ gauge symmetries including U$(1)_Y$.
We examine the anomaly structure of seven U$(1)$ gauge symmetries
 on five D-branes at a singularity (I${}_1$, I${}_2$, I${}_3$).
(The ${\bf Z}_3$ transformation matrices
 for Chan-Paton factors of those D-branes are given in 
 eqs.(\ref{ours-D3}), (\ref{ours-D7}) and (\ref{ours-anti-D7})).
Then, we investigate the mass matrix of whole U$(1)$ gauge bosons.

First, consider mixed non-Abelian anomalies of seven U$(1)$ gauge symmetries
 in D-brane basis:
\begin{equation}
 q_1 = Q_{TC},
\quad
 q_2 \equiv Q_3,
\quad
 q_3 \equiv Q_2,
\quad
 q_4 \equiv Q,
\quad
 q_5 = Q_1',
\quad
 q_6 = Q_2',
\quad
 q_7 = Q_3', 
\end{equation}
 where $Q_{TC}$ is the U$(1)$ charge in U$(3)_{TC}$,
 $Q_i'$ are charges of U$(1)'_i$ from D$7_i$-branes with $i=1,2,3$.
The generators of non-Abelian gauge symmetries are denoted as
 $(T^A)_a$ with $a=1,2,3$ for SU$(3)_{TC}$, SU$(3)_c$ and SU$(2)_L$,
 respectively.
The anomaly matrix is defined as
\begin{equation}
 t_{ia} \equiv {\rm tr} \left[ q_i \left( T^A T^A \right)_a \right]
 = {1 \over 2} \ {\rm tr}_a \left( q_i \right),
\end{equation}
 where ${\rm tr}_a$ means the trace over left-handed fermions
 which belong to fundamental (or anti-fundamental) representation
 under a non-Abelian gauge symmetry of $a$.
Simple calculations give
\begin{equation}
 t_{ia} =
 {1 \over 2}
 \left(
  \begin{array}{ccc}
   0  & 0  & -3 \\
   0  & 0  &  9 \\
   2  & -6 &  0 \\
   -1 & 3  & -3 \\
   -3 & 1  & -1 \\
   -1 & 1  & -1 \\
   -1 & 1  & -1
  \end{array}
 \right).
\end{equation}
It is easily seen that
 there are no mixed non-Abelian anomalies for
 the hypercharge $Y/2 = - (q_2/3 + q_3/2 + q_4)$,
 and a combination $q_7 - q_6$.
These two U$(1)$ gauge symmetries are independent in the sense of
 ${\rm tr}((Y/2)(q_7-q_6))=0$.
There are more two independent U$(1)$ gauge symmetries
 which are free from mixed non-Abelian anomalies,
\begin{eqnarray}
 & 9 q_1 + 3 q_2 - q_4 - q_5 + 2 q_6 + 2 q_7, &
\label{3rd-charge}
\\
 & 49 q_1 + 3 q_2 - 20 q_3 + 9 q_4 + 49 q_5 - 98 q_6 - 98 q_7, &
\label{4th-charge}
\end{eqnarray}
 but these are anomalous with the other U$(1)$'s.

Next, consider the anomalies among U$(1)$'s.
The anomaly matrix is defined as
\begin{equation}
 t_{ijk} \equiv {\rm tr} \left[ q_i q_j q_k \right].
\end{equation}
Since there are no fermion
 which simultaneously has three kinds of U$(1)$ charges,
 it is enough to consider
\begin{equation}
 t_{ij} \equiv t_{ijj}
\end{equation}
Straightforward calculations give
\begin{equation}
 t_{ij}
 =
 \left(
  \begin{array}{ccccccc}
   0  & 0  & -6 & 3  & 9  & 3  & 3  \\
   0  & 0  & 18 & -9 & -3 & -3 & -3 \\
   6  &-18 & 0  & 6  & 2  & 2  & 2  \\
   -3 & 9  & -6 & 0  & 0  & 0  & 0  \\
   -9 & 3  & -2 & 0  & 0  & 0  & 0  \\
   -3 & 3  & -2 & 0  & 0  & 0  & 0  \\
   -3 & 3  & -2 & 0  & 0  & 0  & 0  \\
  \end{array}
 \right).
\end{equation}
Hypercharge and $q_7-q_6$ have no anomalies with any other U$(1)$'s.
The U$(1)$ gauge symmetries
 of eqs.(\ref{3rd-charge}) and (\ref{4th-charge})
 have anomalies with $(q_5, q_6, q_7)$ and $(q_4, q_5, q_6, q_7)$,
 respectively.

The U$(1)$ gauge symmetry of $q_7-q_6$ is anomalous
 with some the other 24 U$(1)$'s on additional D-branes.
Hypercharge is the only U$(1)$ gauge symmetry
 which does not have anomaly with any other U$(1)$'s
 \footnote{
  Since ${\rm tr}(Y/2)=0$, mixed gravitational anomaly is also canceled.}.
Therefore,
 there are no tree-level kinetic term mixings
 between hypercharge and the other anomalous U$(1)$'s
 \footnote{
 The tree-level kinetic term mixing is assumed
  in a phenomenological model in Ref.\cite{Anastasopoulos:2008jt}.
 }.

The masses of the anomalous U$(1)$ gauge bosons in type I theory
 have been explicitly calculated
 in Refs.\cite{Antoniadis:2002cs,Anastasopoulos:2004ga}.
The mass matrix is given in D-brane basis of U$(1)$ gauge bosons.
We translate their results in case of no orientifold projection.
Consider toroidal $(T^2 \times T^2 \times T^2)/{\bf Z}_N$ orbifold
 in type IIB theory and two D-branes at an orbifold singularity
 (two D-branes can be the same one).
The U$(1)$ gauge symmetries on these two D-branes
 are labeled by $a$ and $b$.
There are four cases in D-brane configurations:
\begin{enumerate}
 \item no Dirichlet-Neumann directions with supersymmetry\\
       (D$3$-D$3$ and D$7_i$-D$7_i$, for example).
 \item two pairs of Dirichlet-Neumann directions with supersymmetry\\
       (D$3$-D$7_i$ and D$7_i$-D$7_j$ with $i \ne j$, for example).
 \item no Dirichlet-Neumann directions without supersymmetry\\
       (D$3$-$\overline{{\rm D}3}$ and D$7_i$-$\overline{{\rm D}7_i}$,
        for example).
 \item two pairs of Dirichlet-Neumann directions without supersymmetry\\
       (D$3$-$\overline{{\rm D}7_i}$ and D$7_i$-$\overline{{\rm D}7_j}$
        with $i \ne j$, for example).
\end{enumerate}
The mass formula for each case is given as follows in order.
\begin{eqnarray}
 \alpha' M^2_{ab}
 &=& {{g_a g_b} \over {4 \pi^3 N}} \sum_{k=0}^{N-1}
     \prod_{i=1,2,3} \vert 2 \sin (\pi k v_i) \vert
     {\rm Tr} \left( \left( \gamma_a \right)^k \lambda_a \right)
     \left\{
     {\rm Tr} \left( \left( \gamma_b \right)^k \lambda_b \right)
     \right\}^*,\\
 \alpha' M^2_{ab}
 &=& {{g_a g_b} \over {4 \pi^3 N}} \sum_{k=0}^{N-1}
     \vert 2 \sin (\pi k v_{ab}) \vert
     {\rm Tr} \left( \left( \gamma_a \right)^k \lambda_a \right)
     \left\{
     {\rm Tr} \left( \left( \gamma_b \right)^k \lambda_b \right)
     \right\}^*,\\
 \alpha' M^2_{ab}
 &=& {1 \over 3} \cdot {{g_a g_b} \over {4 \pi^3 N}} \sum_{k=0}^{N-1}
     \prod_{i=1,2,3} \vert 2 \sin (\pi k v_i) \vert
     {\rm Tr} \left( \left( \gamma_a \right)^k \lambda_a \right)
     \left\{
     {\rm Tr} \left( \left( \gamma_b \right)^k \lambda_b \right)
     \right\}^*,\\
 \alpha' M^2_{ab}
 &=& -{1 \over 3} \cdot {{g_a g_b} \over {4 \pi^3 N}} \sum_{k=0}^{N-1}
     \vert 2 \sin (\pi k v_{ab}) \vert
     {\rm Tr} \left( \left( \gamma_a \right)^k \lambda_a \right)
     \left\{
     {\rm Tr} \left( \left( \gamma_b \right)^k \lambda_b \right)
     \right\}^*,\\
\end{eqnarray}
 where $v_i$ is a component of ${\bf Z}_N$ twist vector $v$,
 $v_{ab}$ denotes a component of $v$ corresponding to a pair of
 Dirichlet-Dirichlet or Neumann-Neumann directions,
 $\lambda_a$ and $\gamma_a$ are Chan-Paton factor
 and ${\bf Z}_N$ transformation matrix, respectively,
 and $g_a$ is the gauge coupling of corresponding U$(1)$.
These formulae guarantee that
 the mass matrix is Hermite: $M^2_{ba} = (M^2_{ab})^*$,
 though all the components should be real.

Since writing whole $31 \times 31$ mass matrix here is not meaningful,
 we examine, instead, $7 \times 7$ submatrix for U$(1)$'s
 which we discussed in previous subsection.
\begin{equation}
 \alpha' M^2 \simeq {\sqrt{3} \over {4 \pi^3}}
 \left(
  \begin{array}{ccccccc}
   3 & -{1 \over 3} & {1 \over {3\sqrt{6}}} & {1 \over {6\sqrt{3}}} &
    -{1 \over {2\sqrt{3}}} & {1 \over {6\sqrt{3}}} & {1 \over {6\sqrt{3}}}
   \\
   -{1 \over 3} & 3 & -\sqrt{{3 \over 2}} & -{\sqrt{3} \over 2} &
    -{1 \over {2\sqrt{3}}} & -{1 \over {2\sqrt{3}}} & -{1 \over {2\sqrt{3}}}
   \\
   {1 \over {3\sqrt{6}}} & -\sqrt{{3 \over 2}} & 2 & -{1 \over \sqrt{2}} &
    -{1 \over {3\sqrt{2}}} & -{1 \over {3\sqrt{2}}} & -{1 \over {3\sqrt{2}}}
   \\
   {1 \over {6\sqrt{3}}} & -{\sqrt{3} \over 2} & -{1 \over \sqrt{2}} & 1 &
    {1 \over 3} & {1 \over 3} & {1 \over 3}
   \\
   -{1 \over {2\sqrt{3}}} & -{1 \over {2\sqrt{3}}} & -{1 \over {3\sqrt{2}}} &
    {1 \over 3} & 1 & {1 \over 3} & {1 \over 3}
   \\
   {1 \over {6\sqrt{3}}} & -{1 \over {2\sqrt{3}}} & -{1 \over {3\sqrt{2}}} &
    {1 \over 3} & {1 \over 3} & 1 & {1 \over 3}
   \\
   {1 \over {6\sqrt{3}}} & -{1 \over {2\sqrt{3}}} & -{1 \over {3\sqrt{2}}} &
    {1 \over 3} & {1 \over 3} & {1 \over 3} & 1
  \end{array}
 \right),
\end{equation}
 where we assume for simplicity that
 all the gauge coupling constants are equal and of the order of unity.
There is only one massless eigenstate
 which corresponds to hypercharge U$(1)_Y$.
The U$(1)$ of $q_7 - q_6$ has mass $\sqrt{3} / 6 \pi^3 \alpha'$,
 though it is anomaly free.
This may be understood as the effect of higher dimensional anomalies
 \cite{Ibanez:2001nd,Scrucca:2002is,Antoniadis:2002cs},
 but anyway the gauge boson corresponding to $q_7-q_6$
 is not the mass eigenstate in total 31 U$(1)$'s.
The total $31 \times 31$ mass matrix is calculated in the same way,
 but the analytic forms of its eigenvalues and eigenstates are not simple.
In the limit of small gauge couplings
 on D$7_2$, D$7_3$ and $\overline{{\rm D}7}_2$ branes (large $R$),
 mass eigenstates of 31 gauge bosons are categorized into three groups:
 massless U$(1)_Y$,
 eight massive U$(1)$'s which dominantly couple with ``our world'',
 and 22 massive U$(1)$'s (six of them are very light ($\sim 1$ MeV)
 with very small gauge couplings ($\sim 10^{-7}$))
 which dominantly couple with ``hidden world''.
The gauge bosons of the first and second group
 dominantly consist of U$(1)$'s on D$3$, D$7_1$ and $\overline{{\rm D}7}_1$,
 and the gauge bosons of the third group
 dominantly consist of U$(1)$'s on the other D-branes.
We consider only the second group in the following,
 since the contribution of gauge bosons in the third group to $S$ parameter
 should be very small.

The masses of eight gauge bosons in the second group,
 $X_i$ with $i=1,2,\cdots,8$ are given by
\begin{equation}
 M_{X_i}^2 = {\sqrt{3} \over \pi} (M_s^*)^2 f_i
\end{equation}
 with coefficients $f_i$ numerically calculated as
\begin{eqnarray}
 &
 f_1 \simeq 0.14,
 \quad
 f_2 \simeq 0.97,
 \quad
 f_3 \simeq 1.4,
 \quad
 f_4 \simeq 2.0,
 & \nonumber\\
 &
 f_5 \simeq 2.3,
 \quad
 f_6 \simeq 2.8,
 \quad
 f_7 \simeq 3.4,
 \quad
 f_8 \simeq 4.1.
 & \nonumber
\end{eqnarray}
Since techni-fermion condensates of eq.(\ref{condensates})
 break some U$(1)$ gauge symmetries,
 these masses may receive some corrections of the order of 100 GeV.
In the following we neglect these corrections to the above mass eigenvalues,
 because the masses produced by string effect is larger than 1 TeV
 with the string scale of the order of 10 TeV.
Defining the charge matrices ${\tilde Q}_i$
 numerical calculations give
\begin{eqnarray}
 &
 {\rm tr} \left( Q_{Y/2} \ {\tilde Q}_1 \right) \simeq 2.3,
 \quad
 {\rm tr} \left( Q_{Y/2} \ {\tilde Q}_2 \right) \simeq -0.018,
 \quad
 {\rm tr} \left( Q_{Y/2} \ {\tilde Q}_3 \right) \simeq -2.1,
 & \nonumber\\
 &
 {\rm tr} \left( Q_{Y/2} \ {\tilde Q}_4 \right) \simeq -2.3,
 \quad
 {\rm tr} \left( Q_{Y/2} \ {\tilde Q}_5 \right) \simeq 2.6,
 \quad
 {\rm tr} \left( Q_{Y/2} \ {\tilde Q}_6 \right) \simeq 2.0,
 & \nonumber\\
 &
 {\rm tr} \left( Q_{Y/2} \ {\tilde Q}_7 \right) \simeq -2.7,
 \quad
 {\rm tr} \left( Q_{Y/2} \ {\tilde Q}_8 \right) \simeq 3.2.
 &
\end{eqnarray}
Therefore,
 all eight anomalous U$(1)$ gauge bosons can have kinetic term mixing
 with U$(1)_Y$ gauge boson (and with each other) at one loop level.
We can understand that
 massless non-anomalous U$(1)$ gauge symmetries are not necessary
 independent of massive anomalous U$(1)$ gauge symmetries,
 namely ${\rm tr} ( Q_{\rm non-anomalous} \ Q_{\rm anomalous} ) \ne 0$,
 by explicitly and analytically checking simpler models.

\subsection{Towards a solution of $S$ parameter problem}
\label{S-parameter}

In case there is kinetic-term mixing
 between U$(1)_Y$ gauge boson and a massive vector boson $X$
 without mass mixing between $W_3$ and $X$,
 the contribution to $S$ parameter is given by \cite{Holdom:1990xp}
\begin{equation}
 S = {{16} \over \alpha}
     {{(c^2-(M_X/m_Z)^2) s^2 c^2 \omega^2}
      \over
      {((M_X/m_Z)^2 - 1)^2}},
\end{equation}
 where $\omega$ is the coefficient of the kinetic term mixing
\begin{equation}
 {\cal L}_{\rm kin. mix}
 = \omega F_Y{}^{\mu\nu} F_X{}_{\mu\nu}.
\end{equation}
This is the second order contribution in $\omega$.
The sign is always negative with $M_X > m_Z$,
 which is appropriate to cancel the positive contribution
 of technicolor, eq.(\ref{S-techni}).
Note that heavy $X$, $M_X \gg m_Z$, quickly decouples.
In case $X$ has mass mixing with $W_3$
 there is a contribution of the first order in $\omega$
 through
\begin{equation}
 \Pi_{3Y}'(0) = \Pi_{3X}(0) {{g_X^2} \over {-M_X^2}} \Pi_{YX}'(0)
\end{equation}
 with $\Pi_{YX}'(0) = 2 \omega$ \cite{Kitazawa:1995nb}.
Again, heavy $X$, $M_X \gg \Pi_{3X}(0)$, quickly decouples.

First,
 estimate the values of $\omega$ for eight anomalous U$(1)$'s.
Here, we consider low-energy effective field theory,
 and calculate fermion one-loop diagrams with ultraviolet cutoff.
A simple estimate gives
\begin{equation}
 \omega_i \simeq - {\rm tr} \left( Q_{Y/2} Q_{X_i} \right)
        {{g_Y g_{X_i}} \over {16\pi^2}}
        \ln \left( {{(M_s^*)^2} \over {m^2}} \right)
       \simeq - {\rm tr} \left( Q_{Y/2} Q_{X_i} \right)
                {1 \over {16\pi^2}}
                \ln \left( {{(M_s^*)^2} \over {m^2}} \right),
\label{omega-one-loop}
\end{equation}
 where we set the string scale as ultraviolet cutoff
 and $m^2$ is an infrared cutoff which should be taken as
 fermion masses of the order of 1 GeV.
In addition to this one-loop effect (planer open string one loop),
 there is string effect mediated by closed string
 (non-planer open string one loop).
Some explicit calculations in String Theory have already been done
 in Refs.\cite{Abel:2003ue,Abel:2008ai},
 but the subtraction of divergences due to NS-NS tadpoles
 (here, NS is abbreviation of Neveu-Schwarz) makes results ambiguous.
For example,
 we can read from their results
 the coefficient of kinetic term mixing
 between D$3$ and $\overline{{\rm D}7}_1$ branes as
\begin{equation}
 \omega \simeq
 {\rm tr} \left( \lambda_{D3} \right) 
 {\rm tr} \left( \lambda_{\overline{D7}} \right)
 {4 \over {(2 \pi)^3}}
 \sum_{n_1,n_2 \ne 0} {1 \over {n_1^2 + n_2^2}},
\end{equation}
 where summations are taken over winding modes
 in two Dirichlet-Dirichlet directions.
The summation diverges due to NS-NS tadpoles,
 and the authors in Ref.\cite{Abel:2003ue} take the contribution
 from the first winding modes only and obtain a finite result.
The procedure of tadpole resummations
 \cite{Dudas:2004nd,Kitazawa:2008hv,Kitazawa:2008tb}
 may be appropriate to obtain unambiguous results.
The true value
 could be larger than the naive estimation of eq.(\ref{omega-one-loop}).
Note that there are ambiguities in $\omega$,
 and we leave this problem for future works.

The second order contribution in the coefficient of kinetic term mixing
 between hypercharge gauge boson and lightest anomalous U$(1)$
 gauge boson to $S$ parameter is
\begin{equation}
  S \simeq -0.03,
\end{equation}
 where we do not consider kinetic term mixings
 among eight anomalous U$(1)$'s, for simplicity.
Though the sign is appropriate to cancel the large contribution
 by technicolor dynamics, $S_{TC} \simeq 0.32$,
 the absolute value is one order smaller.
The contribution of next lightest anomalous U(1)
 is at least two order smaller than the above value.
We need a little lighter anomalous U$(1)$ and/or
 a little larger kinetic term mixing.
For example,
 factor three large $\omega$ is enough
 to satisfy the experimental constraint.

The first order contributions depend on mass mixings
 between $W_3$ and anomalous U$(1)$'s through electroweak symmetry breaking.
The mass mixing is determined by the charges of technicolor condensates.
\begin{equation}
 m_{3X_i}^2 = m_{X_i3}^2
 \simeq {\rm (100 \ GeV)}^2 \Delta Q_{X_i}
\end{equation}
 with $\Delta Q_{X_i} \equiv
 \left[ Q_{X_i}(\Psi_L^{I_3=+1/2} \Psi_{7_2})
        - Q_{X_i}(\Psi_L^{I_3=-1/2} \Psi_E) \right]$,
 where
\begin{eqnarray}
&
 \Delta Q_{X_1} = -0.087,
\quad
 \Delta Q_{X_2} = -0.011,
\quad
 \Delta Q_{X_3} = 0.15,
\quad
 \Delta Q_{X_4} = 0.18,
&
\nonumber\\
&
 \Delta Q_{X_5} = -0.38,
\quad
 \Delta Q_{X_6} = -0.11,
\quad
 \Delta Q_{X_7} = 0.019,
\quad
 \Delta Q_{X_i} = -0.095.
&
\end{eqnarray}
These values of the order of $1/10$ reflect that
 each mass eigenstate has about $10\%$ component
 of ``hidden'' anomalous U$(1)$'s.
The contribution to $S$ parameter is
\begin{equation}
 S \simeq
  - 16 \pi \sum_{i=1}^8 \left( 2 m_{3X_i}^2
                               \cdot {1 \over {-M_{X_i}^2}}
                               \cdot 2 \omega_{X_i} \right)
   \simeq 0.01.
\end{equation}
This is the same order of magnitude of the second order contribution
 with opposite sign.
The first order contribution
 is not larger than the second order contribution,
 since the masses by electroweak symmetry breaking
 are much smaller than the original masses of anomalous U$(1)$'s.
The sign of this contribution is very model dependent.

\section{Conclusions}
\label{conclusions}

We have examined
 the possibility of dynamical electroweak symmetry breaking (technicolor)
 in string models
 by concretely analyzing a toy model
 mainly concentrating on the solution of $S$ parameter problem
 with massive anomalous U$(1)$ gauge bosons.
It has been found that
 the contributions of massive anomalous U$(1)$'s to $S$ parameter
 are non-negligible,
 and they even have potential to cancel large technicolor contribution.
Since there is no general relation between the magnitudes of
 technicolor and anomalous U$(1)$ contributions to $S$ parameter,
 the cancellation, which may happen, is accidental.

It is very likely that
 anomalous U$(1)$ gauge bosons give large contribution to $T$ parameter
 as well as $S$ parameter \cite{Ghilencea:2002da}.
It is known that
 $T$ parameter is sensitive to
 the mechanism of quark and lepton mass generations,
 especially the generation of top quark mass (or top-bottom mass splitting).
The string scale of the order of 10 TeV is appropriate to generate mass
 of the order of $2 \pi (100)^3 / (10000)^2 \simeq 0.06$ GeV,
 and it is apparently difficult to generate large top quark mass.
This is more serious problem in technicolor scenario in general
 than $S$ parameter problem.
It would be interesting to pursue the solution of this problem
 in string models beyond the framework of field theory.

There is a tension
 between dynamical electroweak symmetry breaking
 and light anomalous U$(1)$ gauge bosons.
To have light, of the order of TeV, anomalous U$(1)$ gauge bosons,
 the string scale should be less than 10 TeV.
We have seen in our toy model that
 we need large threshold correction to the technicolor gauge coupling
 at the string scale so that it becomes strong at the weak scale.
Larger threshold correction is required for smaller string scale.
In the theoretical point of view, this scenario is very constrained.
Near future collider experiments and astronomical observations
 will give strong constraints to this scenario.

The landscape analysis requiring additional non-Abelian gauge symmetry
 for dynamical electroweak symmetry breaking
 (without elementary Higgs doublet fields)
 would be interesting in the theoretical point of view.

\section*{Acknowledgments}

I thank the Galileo Galilei Institute for Theoretical Physics
 for the hospitality and the INFN for partial support
 during the early stage of this work.
I would like to thank Augusto Sagnotti
 for providing chance and support to join the workshop
 ``New Perspectives in String Theory'' at that institute.
It is a great pleasure to thank
 Ignatios Antoniadis, Massimo Bianchi and Emilian Dudas
 for helpful discussions.
I would like to thank Emilian Dudas for comments on the paper.
This work was supported in part by INFN,
 and in part by the Italian MIUR-PRIN contract 2007-5ATT78.


\begin{thebibliography}{99}

\bibitem{Haber:1993wf}
  H.~E.~Haber,
  ``Introductory low-energy supersymmetry,''
  arXiv:hep-ph/9306207.
\bibitem{Nilles:1995ci}
  H.~P.~Nilles,
  ``Phenomenological aspects of supersymmetry,''
  arXiv:hep-ph/9511313.
\bibitem{Baer:1995tb}
  H.~Baer {\it et al.},
  ``Low-energy supersymmetry phenomenology,''
  arXiv:hep-ph/9503479.
\bibitem{Bagger:1996ka}
  J.~A.~Bagger,
  ``Weak-scale supersymmetry: Theory and practice,''
  arXiv:hep-ph/9604232.
\bibitem{Castano:1993ri}
  D.~J.~Castano, E.~J.~Piard and P.~Ramond,
  ``Renormalization Group Study Of The Standard Model And Its Extensions. 2.
  The Minimal Supersymmetric Standard Model,''
  Phys.\ Rev.\  D {\bf 49} (1994) 4882
  [arXiv:hep-ph/9308335].
\bibitem{Drees:1995hj}
  M.~Drees and S.~P.~Martin,
  ``Implications of SUSY model building,''
  arXiv:hep-ph/9504324.
\bibitem{Inoue:1982pi}
  K.~Inoue, A.~Kakuto, H.~Komatsu and S.~Takeshita,
  ``Aspects Of Grand Unified Models With Softly Broken Supersymmetry,''
  Prog.\ Theor.\ Phys.\  {\bf 68} (1982) 927
  [Erratum-ibid.\  {\bf 70} (1983) 330].
\bibitem{ArkaniHamed:2001nc}
  N.~Arkani-Hamed, A.~G.~Cohen and H.~Georgi,
  ``Electroweak symmetry breaking from dimensional deconstruction,''
  Phys.\ Lett.\  B {\bf 513} (2001) 232
  [arXiv:hep-ph/0105239].
\bibitem{ArkaniHamed:2002qy}
  N.~Arkani-Hamed, A.~G.~Cohen, E.~Katz and A.~E.~Nelson,
  ``The littlest Higgs,''
  JHEP {\bf 0207} (2002) 034
  [arXiv:hep-ph/0206021].
\bibitem{Antoniadis:1990ew}
  I.~Antoniadis,
  ``A Possible new dimension at a few TeV,''
  Phys.\ Lett.\  B {\bf 246} (1990) 377.
\bibitem{ArkaniHamed:1998rs}
  N.~Arkani-Hamed, S.~Dimopoulos and G.~R.~Dvali,
  ``The hierarchy problem and new dimensions at a millimeter,''
  Phys.\ Lett.\  B {\bf 429} (1998) 263
  [arXiv:hep-ph/9803315].
\bibitem{Randall:1999ee}
  L.~Randall and R.~Sundrum,
  ``A large mass hierarchy from a small extra dimension,''
  Phys.\ Rev.\ Lett.\  {\bf 83} (1999) 3370
  [arXiv:hep-ph/9905221].
\bibitem{Fairlie:1979at}
  D.~B.~Fairlie,
  ``Higgs' Fields And The Determination Of The Weinberg Angle,''
  Phys.\ Lett.\  B {\bf 82} (1979) 97.
\bibitem{Manton:1979kb}
  N.~S.~Manton,
  ``A New Six-Dimensional Approach To The Weinberg-Salam Model,''
  Nucl.\ Phys.\  B {\bf 158} (1979) 141.
\bibitem{Hosotani:1983xw}
  Y.~Hosotani,
  ``Dynamical Mass Generation By Compact Extra Dimensions,''
  Phys.\ Lett.\  B {\bf 126} (1983) 309.
\bibitem{Csaki:2003dt}
  C.~Csaki, C.~Grojean, H.~Murayama, L.~Pilo and J.~Terning,
  ``Gauge theories on an interval: Unitarity without a Higgs,''
  Phys.\ Rev.\  D {\bf 69} (2004) 055006
  [arXiv:hep-ph/0305237].
\bibitem{Weinberg:1975gm}
  S.~Weinberg,
  ``Implications Of Dynamical Symmetry Breaking,''
  Phys.\ Rev.\  D {\bf 13} (1976) 974.
\bibitem{Weinberg:1979bn}
  S.~Weinberg,
  ``Implications Of Dynamical Symmetry Breaking: An Addendum,''
  Phys.\ Rev.\  D {\bf 19} (1979) 1277.
\bibitem{Susskind:1978ms}
  L.~Susskind,
  ``Dynamics Of Spontaneous Symmetry Breaking
   In The Weinberg-Salam Theory,''
  Phys.\ Rev.\  D {\bf 20} (1979) 2619.
\bibitem{Miransky:1988xi}
  V.~A.~Miransky, M.~Tanabashi and K.~Yamawaki,
  ``Dynamical Electroweak Symmetry Breaking
   with Large Anomalous Dimension and t Quark Condensate,''
  Phys.\ Lett.\  B {\bf 221} (1989) 177.
\bibitem{Nambu:1989jt}
  Y.~Nambu,
  ``BOOTSTRAP SYMMETRY BREAKING IN ELECTROWEAK UNIFICATION,''
  EFI-89-08.
\bibitem{Marciano:1989xd}
  W.~J.~Marciano,
  ``HEAVY TOP QUARK MASS PREDICTIONS,''
  Phys.\ Rev.\ Lett.\  {\bf 62} (1989) 2793.
\bibitem{Bardeen:1989ds}
  W.~A.~Bardeen, C.~T.~Hill and M.~Lindner,
  ``Minimal Dynamical Symmetry Breaking Of The Standard Model,''
  Phys.\ Rev.\  D {\bf 41} (1990) 1647.
\bibitem{Belyaev:2008yj}
  A.~Belyaev, R.~Foadi, M.~T.~Frandsen, M.~Jarvinen,
   F.~Sannino and A.~Pukhov,
  ``Technicolor Walks at the LHC,''
  Phys.\ Rev.\  D {\bf 79} (2009) 035006
  [arXiv:0809.0793 [hep-ph]].
\bibitem{Giddings:2001yu}
  S.~B.~Giddings, S.~Kachru and J.~Polchinski,
  ``Hierarchies from fluxes in string compactifications,''
  Phys.\ Rev.\  D {\bf 66} (2002) 106006
  [arXiv:hep-th/0105097].
\bibitem{Aldazabal:2000dg}
  G.~Aldazabal, S.~Franco, L.~E.~Ibanez, R.~Rabadan and A.~M.~Uranga,
  ``D = 4 chiral string compactifications from intersecting branes,''
  J.\ Math.\ Phys.\  {\bf 42} (2001) 3103
  [arXiv:hep-th/0011073].
\bibitem{Cremades:2002cs}
  D.~Cremades, L.~E.~Ibanez and F.~Marchesano,
  ``Intersecting brane models of particle physics and the Higgs mechanism,''
  JHEP {\bf 0207} (2002) 022
  [arXiv:hep-th/0203160].
\bibitem{Antoniadis:2000tq}
  I.~Antoniadis, K.~Benakli and M.~Quiros,
  ``Radiative symmetry breaking in brane models,''
  Nucl.\ Phys.\  B {\bf 583} (2000) 35
  [arXiv:hep-ph/0004091].
\bibitem{Kitazawa:2006if}
  N.~Kitazawa,
  ``Radiative symmetry breaking on D-branes
   at non-supersymmetric singularities,''
  Nucl.\ Phys.\  B {\bf 755} (2006) 254
  [arXiv:hep-th/0606182].
\bibitem{Bachas:1995ik}
  C.~Bachas,
  ``A Way to break supersymmetry,''
  arXiv:hep-th/9503030.
\bibitem{Blumenhagen:2000wh}
  R.~Blumenhagen, L.~Goerlich, B.~Kors and D.~Lust,
  ``Noncommutative compactifications of type I strings on tori
   with  magnetic background flux,''
  JHEP {\bf 0010} (2000) 006
  [arXiv:hep-th/0007024].
\bibitem{Angelantonj:2000hi}
  C.~Angelantonj, I.~Antoniadis, E.~Dudas and A.~Sagnotti,
  ``Type-I strings on magnetised orbifolds and brane transmutation,''
  Phys.\ Lett.\  B {\bf 489} (2000) 223
  [arXiv:hep-th/0007090].
\bibitem{Aldazabal:2000cn}
  G.~Aldazabal, S.~Franco, L.~E.~Ibanez, R.~Rabadan and A.~M.~Uranga,
  ``Intersecting brane worlds,''
  JHEP {\bf 0102} (2001) 047
  [arXiv:hep-ph/0011132].
\bibitem{Ibanez:2001nd}
  L.~E.~Ibanez, F.~Marchesano and R.~Rabadan,
  ``Getting just the standard model at intersecting branes,''
  JHEP {\bf 0111} (2001) 002
  [arXiv:hep-th/0105155].
\bibitem{Blumenhagen:2001te}
  R.~Blumenhagen, B.~Kors, D.~Lust and T.~Ott,
  ``The standard model from stable intersecting brane world orbifolds,''
  Nucl.\ Phys.\  B {\bf 616} (2001) 3
  [arXiv:hep-th/0107138].
\bibitem{Cvetic:2001tj}
  M.~Cvetic, G.~Shiu and A.~M.~Uranga,
  ``Three-family supersymmetric standard like models
   from intersecting brane worlds,''
  Phys.\ Rev.\ Lett.\  {\bf 87} (2001) 201801
  [arXiv:hep-th/0107143].
\bibitem{Cvetic:2001nr}
  M.~Cvetic, G.~Shiu and A.~M.~Uranga,
  ``Chiral four-dimensional N = 1 supersymmetric type IIA orientifolds
   from intersecting D6-branes,''
  Nucl.\ Phys.\  B {\bf 615} (2001) 3
  [arXiv:hep-th/0107166].
\bibitem{Kokorelis:2002zz}
  C.~Kokorelis,
  ``New standard model vacua from intersecting branes,''
  JHEP {\bf 0209} (2002) 029
  [arXiv:hep-th/0205147].
\bibitem{Aldazabal:2000sa}
  G.~Aldazabal, L.~E.~Ibanez, F.~Quevedo and A.~M.~Uranga,
  ``D-branes at singularities:
   A bottom-up approach to the string embedding of the standard model,''
  JHEP {\bf 0008} (2000) 002
  [arXiv:hep-th/0005067].
\bibitem{Higaki:2005ie}
  T.~Higaki, N.~Kitazawa, T.~Kobayashi and K.~j.~Takahashi,
  ``Flavor structure and coupling selection rule
   from intersecting  D-branes,''
  Phys.\ Rev.\  D {\bf 72} (2005) 086003
  [arXiv:hep-th/0504019].
\bibitem{Anastasopoulos:2006da}
  P.~Anastasopoulos, T.~P.~T.~Dijkstra, E.~Kiritsis and A.~N.~Schellekens,
  ``Orientifolds, hypercharge embeddings and the standard model,''
  Nucl.\ Phys.\  B {\bf 759} (2006) 83
  [arXiv:hep-th/0605226].
\bibitem{Kitazawa:2004ed}
  N.~Kitazawa,
  ``Supersymmetric composite models on intersecting D-branes,''
  Nucl.\ Phys.\  B {\bf 699} (2004) 124
  [arXiv:hep-th/0401096].
\bibitem{Kitazawa:2004hz}
  N.~Kitazawa,
  ``Dynamical generation of mu-terms and Yukawa couplings
   in intersecting D-brane models,''
  JHEP {\bf 0411} (2004) 044
  [arXiv:hep-th/0403278].
\bibitem{Kitazawa:2004nf}
  N.~Kitazawa, T.~Kobayashi, N.~Maru and N.~Okada,
  ``Yukawa coupling structure in intersecting D-brane models,''
  Eur.\ Phys.\ J.\  C {\bf 40} (2005) 579
  [arXiv:hep-th/0406115].
\bibitem{Antoniadis:2000jv}
  I.~Antoniadis, K.~Benakli and A.~Laugier,
  ``Contact interactions in D-brane models,''
  JHEP {\bf 0105} (2001) 044
  [arXiv:hep-th/0011281].
\bibitem{Abel:2004rp}
  S.~Abel and J.~Santiago,
  ``Constraining the string scale: from Planck to Weak and back again,''
  J.\ Phys.\ G {\bf 30} (2004) R83
  [arXiv:hep-ph/0404237].
\bibitem{Kitazawa:2008tb}
  N.~Kitazawa,
  ``One-loop masses of open-string scalar fields in String Theory,''
  JHEP {\bf 0809} (2008) 049
  [arXiv:0805.0824 [hep-th]].
\bibitem{Conlon:2008wa}
  J.~P.~Conlon, A.~Maharana and F.~Quevedo,
  ``Towards Realistic String Vacua,''
  JHEP {\bf 0905} (2009) 109
  [arXiv:0810.5660 [hep-th]].
\bibitem{Sundrum:1991rf}
  R.~Sundrum and S.~D.~H.~Hsu,
  ``Walking technicolor and electroweak radiative corrections,''
  Nucl.\ Phys.\  B {\bf 391} (1993) 127
  [arXiv:hep-ph/9206225].
\bibitem{Appelquist:1998xf}
  T.~Appelquist and F.~Sannino,
  ``The Physical Spectrum of Conformal SU(N) Gauge Theories,''
  Phys.\ Rev.\  D {\bf 59} (1999) 067702
  [arXiv:hep-ph/9806409].
\bibitem{Hong:2006si}
  D.~K.~Hong and H.~U.~Yee,
  ``Holographic estimate of oblique corrections for technicolor,''
  Phys.\ Rev.\  D {\bf 74} (2006) 015011
  [arXiv:hep-ph/0602177].
\bibitem{Holdom:1990xp}
  B.~Holdom,
  ``Oblique electroweak corrections and an extra gauge boson,''
  Phys.\ Lett.\  B {\bf 259} (1991) 329.
\bibitem{Kitazawa:1995nb}
  N.~Kitazawa and T.~Yanagida,
  ``A viable one-family technicolor model,''
  Phys.\ Lett.\  B {\bf 383} (1996) 78
  [arXiv:hep-ph/9510228].
\bibitem{Sugimoto:1999tx}
  S.~Sugimoto,
  ``Anomaly cancellations in type I D9-D9-bar system and the USp(32)
    string theory,''
  Prog.\ Theor.\ Phys.\  {\bf 102} (1999) 685
  [arXiv:hep-th/9905159].
\bibitem{Antoniadis:1999xk}
  I.~Antoniadis, E.~Dudas and A.~Sagnotti,
  ``Brane supersymmetry breaking,''
  Phys.\ Lett.\  B {\bf 464} (1999) 38
  [arXiv:hep-th/9908023].
\bibitem{Angelantonj:1999jh}
  C.~Angelantonj,
  ``Comments on open-string orbifolds with a non-vanishing B(ab),''
  Nucl.\ Phys.\  B {\bf 566} (2000) 126
  [arXiv:hep-th/9908064].
\bibitem{Aldazabal:1999jr}
  G.~Aldazabal and A.~M.~Uranga,
  ``Tachyon-free non-supersymmetric type IIB orientifolds
    via brane-antibrane systems,''
  JHEP {\bf 9910} (1999) 024
  [arXiv:hep-th/9908072].
\bibitem{Angelantonj:1999ms}
  C.~Angelantonj, I.~Antoniadis, G.~D'Appollonio, E.~Dudas and A.~Sagnotti,
  ``Type I vacua with brane supersymmetry breaking,''
  Nucl.\ Phys.\  B {\bf 572} (2000) 36
  [arXiv:hep-th/9911081].
\bibitem{Antoniadis:2004qn}
  I.~Antoniadis and T.~R.~Taylor,
  ``Topological masses from broken supersymmetry,''
  Nucl.\ Phys.\  B {\bf 695} (2004) 103
  [arXiv:hep-th/0403293].
\bibitem{Antoniadis:2005sd}
  I.~Antoniadis, K.~S.~Narain and T.~R.~Taylor,
  ``Open string topological amplitudes and gaugino masses,''
  Nucl.\ Phys.\  B {\bf 729} (2005) 235
  [arXiv:hep-th/0507244].
\bibitem{Carone:2007md}
  C.~D.~Carone, J.~Erlich and M.~Sher,
  ``Holographic Electroweak Symmetry Breaking from D-branes,''
  Phys.\ Rev.\  D {\bf 76} (2007) 015015
  [arXiv:0704.3084 [hep-th]].
\bibitem{Farhi:1980xs}
  E.~Farhi and L.~Susskind,
  ``Technicolor,''
  Phys.\ Rept.\  {\bf 74} (1981) 277.
\bibitem{Peskin:1990zt}
  M.~E.~Peskin and T.~Takeuchi,
  ``A New constraint on a strongly interacting Higgs sector,''
  Phys.\ Rev.\ Lett.\  {\bf 65} (1990) 964.
\bibitem{Peskin:1991sw}
  M.~E.~Peskin and T.~Takeuchi,
  ``Estimation of oblique electroweak corrections,''
  Phys.\ Rev.\  D {\bf 46} (1992) 381.
\bibitem{Altarelli:1990zd}
  G.~Altarelli and R.~Barbieri,
  ``Vacuum polarization effects of new physics on electroweak processes,''
  Phys.\ Lett.\  B {\bf 253} (1991) 161.
\bibitem{Altarelli:1991fk}
  G.~Altarelli, R.~Barbieri and S.~Jadach,
  ``Toward a model independent analysis of electroweak data,''
  Nucl.\ Phys.\  B {\bf 369} (1992) 3
  [Erratum-ibid.\  B {\bf 376} (1992) 444].
\bibitem{Amsler:2008zzb}
  C.~Amsler {\it et al.}  [Particle Data Group],
  ``Review of particle physics,''
  Phys.\ Lett.\  B {\bf 667} (2008) 1.
\bibitem{Sannino:2004qp}
  F.~Sannino and K.~Tuominen,
  ``Orientifold theory dynamics and symmetry breaking,''
  Phys.\ Rev.\  D {\bf 71} (2005) 051901
  [arXiv:hep-ph/0405209].
\bibitem{Dietrich:2005jn}
  D.~D.~Dietrich, F.~Sannino and K.~Tuominen,
  ``Light composite Higgs from higher representations versus electroweak
  precision measurements: Predictions for LHC,''
  Phys.\ Rev.\  D {\bf 72} (2005) 055001
  [arXiv:hep-ph/0505059].
\bibitem{Foadi:2007ue}
  R.~Foadi, M.~T.~Frandsen, T.~A.~Ryttov and F.~Sannino,
  ``Minimal Walking Technicolor: Set Up for Collider Physics,''
  Phys.\ Rev.\  D {\bf 76} (2007) 055005
  [arXiv:0706.1696 [hep-ph]].
\bibitem{Hanhart:2001fx}
  C.~Hanhart, J.~A.~Pons, D.~R.~Phillips and S.~Reddy,
  ``The likelihood of GODs' existence:
   Improving the SN1987a constraint on the size of large compact dimensions,''
  Phys.\ Lett.\  B {\bf 509} (2001) 1
  [arXiv:astro-ph/0102063].
\bibitem{Hannestad:2001xi}
  S.~Hannestad and G.~G.~Raffelt,
  ``Stringent neutron-star limits on large extra dimensions,''
  Phys.\ Rev.\ Lett.\  {\bf 88} (2002) 071301
  [arXiv:hep-ph/0110067].
\bibitem{Hannestad:2003yd}
  S.~Hannestad and G.~G.~Raffelt,
  ``Supernova and neutron-star limits on large extra dimensions reexamined,''
  Phys.\ Rev.\  D {\bf 67} (2003) 125008
  [Erratum-ibid.\  D {\bf 69} (2004) 029901]
  [arXiv:hep-ph/0304029].
\bibitem{Antoniadis:1999ge}
  I.~Antoniadis, C.~Bachas and E.~Dudas,
  ``Gauge couplings in four-dimensional type I string orbifolds,''
  Nucl.\ Phys.\  B {\bf 560} (1999) 93
  [arXiv:hep-th/9906039].
\bibitem{Bachas:1996zt}
  C.~Bachas and C.~Fabre,
  ``Threshold Effects in Open-String Theory,''
  Nucl.\ Phys.\  B {\bf 476} (1996) 418
  [arXiv:hep-th/9605028].
\bibitem{Bachas:1998kr}
  C.~P.~Bachas,
  ``Unification with low string scale,''
  JHEP {\bf 9811} (1998) 023
  [arXiv:hep-ph/9807415].
\bibitem{Ghilencea:2002ff}
  D.~M.~Ghilencea and S.~Groot Nibbelink,
  ``String threshold corrections from field theory,''
  Nucl.\ Phys.\  B {\bf 641} (2002) 35
  [arXiv:hep-th/0204094].
\bibitem{Anastasopoulos:2006hn}
  P.~Anastasopoulos, M.~Bianchi, G.~Sarkissian and Y.~S.~Stanev,
  ``On gauge couplings and thresholds in type I gepner models and otherwise,''
  JHEP {\bf 0703} (2007) 059
  [arXiv:hep-th/0612234].
\bibitem{Lust:2003ky}
  D.~Lust and S.~Stieberger,
  ``Gauge threshold corrections in intersecting brane world models,''
  Fortsch.\ Phys.\  {\bf 55} (2007) 427
  [arXiv:hep-th/0302221].
\bibitem{Akerblom:2007np}
  N.~Akerblom, R.~Blumenhagen, D.~Lust and M.~Schmidt-Sommerfeld,
  ``Thresholds for intersecting D-branes revisited,''
  Phys.\ Lett.\  B {\bf 652} (2007) 53
  [arXiv:0705.2150 [hep-th]].
\bibitem{Benakli:2008ub}
  K.~Benakli and M.~D.~Goodsell,
  ``Two-Point Functions of Chiral Fields at One Loop in Type II,''
  Nucl.\ Phys.\  B {\bf 805} (2008) 72
  [arXiv:0805.1874 [hep-th]].
\bibitem{Conlon:2009xf}
  J.~P.~Conlon,
  ``Gauge Threshold Corrections for Local String Models,''
  JHEP {\bf 0904} (2009) 059
  [arXiv:0901.4350 [hep-th]].
\bibitem{Conlon:2009kt}
  J.~P.~Conlon and E.~Palti,
  ``Gauge Threshold Corrections for Local Orientifolds,''
  arXiv:0906.1920 [hep-th].
\bibitem{Anastasopoulos:2008jt}
  P.~Anastasopoulos, F.~Fucito, A.~Lionetto, G.~Pradisi, A.~Racioppi
   and Y.~S.~Stanev,
  ``Minimal Anomalous U(1) -prime Extension of the MSSM,''
  Phys.\ Rev.\  D {\bf 78} (2008) 085014
  [arXiv:0804.1156 [hep-th]].
\bibitem{Antoniadis:2002cs}
  I.~Antoniadis, E.~Kiritsis and J.~Rizos,
  ``Anomalous U(1)s in type I superstring vacua,''
  Nucl.\ Phys.\  B {\bf 637} (2002) 92
  [arXiv:hep-th/0204153].
\bibitem{Anastasopoulos:2004ga}
  P.~Anastasopoulos,
  ``Anomalous U(1)s masses in non-supersymmetric open string vacua,''
  Phys.\ Lett.\  B {\bf 588} (2004) 119
  [arXiv:hep-th/0402105].
\bibitem{Scrucca:2002is}
  C.~A.~Scrucca, M.~Serone and M.~Trapletti,
  ``Open string models with Scherk-Schwarz SUSY breaking
   and localized anomalies,''
  Nucl.\ Phys.\  B {\bf 635} (2002) 33
  [arXiv:hep-th/0203190].
\bibitem{Abel:2003ue}
  S.~A.~Abel and B.~W.~Schofield,
  ``Brane-antibrane kinetic mixing, millicharged particles and SUSY
  breaking,''
  Nucl.\ Phys.\  B {\bf 685} (2004) 150
  [arXiv:hep-th/0311051].
\bibitem{Abel:2008ai}
  S.~A.~Abel, M.~D.~Goodsell, J.~Jaeckel, V.~V.~Khoze and A.~Ringwald,
  ``Kinetic Mixing of the Photon with Hidden U(1)s in String Phenomenology,''
  JHEP {\bf 0807} (2008) 124
  [arXiv:0803.1449 [hep-ph]].
\bibitem{Dudas:2004nd}
  E.~Dudas, G.~Pradisi, M.~Nicolosi and A.~Sagnotti,
  ``On tadpoles and vacuum redefinitions in string theory,''
  Nucl.\ Phys.\  B {\bf 708} (2005) 3
  [arXiv:hep-th/0410101].
\bibitem{Kitazawa:2008hv}
  N.~Kitazawa,
  ``Tadpole Resummations in String Theory,''
  Phys.\ Lett.\  B {\bf 660} (2008) 415
  [arXiv:0801.1702 [hep-th]].
\bibitem{Ghilencea:2002da}
  D.~M.~Ghilencea, L.~E.~Ibanez, N.~Irges and F.~Quevedo,
  ``TeV-Scale Z' Bosons from D-branes,''
  JHEP {\bf 0208} (2002) 016
  [arXiv:hep-ph/0205083].
\end{thebibliography}
\end{document}